\documentclass[iop]{emulateapj}

\newcommand{\msun}{\ensuremath{\mathit{\,M_\odot}}}

\newcommand{\kps}{\ensuremath{\mathrm{\,km\,s^{-1}}}}
\newcommand{\va}{\ensuremath{\mathrm{V_A}}}
\newcommand{\vb}{\ensuremath{\mathrm{V_B}}}

\tabletypesize{\normalsize}

\shorttitle{Spectroscopic Dynamical Masses of Kepler-16A \& B}
\shortauthors{Bender et al.}

\journalinfo{Accepted to the Astrophysical Journal Letters}
\submitted{}
\begin{document}

\title{The SDSS--HET Survey of \emph{Kepler} Eclipsing Binaries: Spectroscopic Dynamical Masses of the Kepler-16 Circumbinary Planet Hosts\footnotemark[1]}

\footnotetext[1]{Based on observations obtained with the Hobby-Eberly Telescope, which is a joint project of the University of Texas at Austin, the Pennsylvania State University, Stanford University, Ludwig-Maximilians-Universit\"{a}t M\"{u}nchen, and Georg-August-Universit\"{a}t G\"{o}ttingen.}

\author{Chad F.\ Bender, Suvrath Mahadevan, Rohit Deshpande, Jason
  T.\ Wright, Arpita Roy, Ryan C.\ Terrien, Steinn Sigurdsson,
  Lawrence W.\ Ramsey, Donald P.\ Schneider, and Scott W.\ Fleming} 
\affil{Department of Astronomy \& Astrophysics, Pennsylvania State University, 525 Davey Lab, University Park, PA-16802\\Center for Exoplanets \& Habitable
  Worlds, Pennsylvania State University} 
\email{cbender@psu.edu}


\begin{abstract}
We have used high-resolution spectroscopy to observe the Kepler-16
eclipsing binary as a double-lined system, and measure precise radial
velocities for both stellar components.  These velocities yield a
dynamical mass-ratio of $q=0.2994\pm0.0031$.  When combined with the
inclination, $i=90^{\circ}\!\!.3401^{+0.0016}_{-0.0019}$, measured from
the Kepler photometric data by \citet{Doyle11}, we derive dynamical
masses for the Kepler-16 components of
$\mathrm{M_A=0.654\pm0.017\msun}$ and
$\mathrm{M_B=0.1959\pm0.0031\msun}$, a precision of 2.5\% and 1.5\%
respectively.  Our results confirm at the $\sim$2\% level the
mass-ratio derived by D11 with their photometric-dynamical model,
$q=0.2937\pm0.0006$.  These are among the most precise spectroscopic
dynamical masses ever measured for low-mass stars, and provide an
important direct test of the results from the photometric-dynamical
modeling technique.
\end{abstract}


\section{Introduction}

The NASA {\it Kepler} space mission is monitoring about 150,000 stars
with a photometric precision of a few parts per million in an effort
to find transiting Earth mass planets \citep{Borucki10}.  This program
has lead to the discovery of more than 2000 eclipsing binaries (EBs)
in the {\it Kepler} field, for which public light-curves now exist
\citep{Prsa11,Slawson11}.  These light-curves constrain the
inclination angle, $i$, of the orbit, and when coupled with radial
velocities (RVs) derived from double-lined spectroscopic observations
(SB2), they yield precise dynamical masses. Detailed modeling of the
light-curves also provides radii measurements for many of these
targets.  Here, we investigate the properties of the EB Kepler-16A \&
B \citep[][hereafter, D11]{Doyle11}, which hosts a circumbinary planet
Kepler-16b. D11 used a photometric-dynamical model (hereafter, PDM) to
solve the \emph{Kepler} light-curve data for the Newtonian motion of
the three body system, and derived precise masses ($\lesssim 0.5\%$
for the stars, and $\lesssim 5\%$ for the planet) and radii
($\sim0.3\%$ for all bodies).  Our objective here is to use the
traditional RV based approach to independently measure the masses for
Kepler-16A \& B.

Fundamental measurements of a diverse sample of stars are critical for
testing theoretical stellar mass-radius relationships, but placing
meaningful constraints on models requires measurement precision of
better than 2-3\% \citep{Torres10}.  The
DEBCat\footnote{http://www.astro.keele.ac.uk/jkt/debcat/} catalog
lists the physical properties of detached EBs with mass and radius
known to better than $\sim2\%$, following the criteria of
\citet{Andersen91}, and currently includes about 130 such systems.
Most DEBCat stars are solar mass or greater, and models and
observations agree in this regime.  Constraints on the low-mass
population are much sparser: DEBCat includes only 28 stars with
$M<0.8\msun$, and only 3 with $M<0.3\msun$. Many of these have radii
10-20\% larger than predicted by models \citep[e.g.,][]{Lopez07a,
  Lopez07b,Torres10}.  Current theories propose that increased
magnetic fields due to tidal locking produce starspots, resulting in
decreased effective temperature.  In addition, the magnetic fields can
suppress convection in the outer atmosphere of the star. Both
processes can yield inflated stellar radii relative to current models
\citep{Ribas06,Chabrier07}.

Adequately testing these theories requires expanding the sample of
low-mass stars with precisely measured masses and radii, and
spectroscopic observations of the {\it Kepler} EBs as SB2s can yield
such a sample.  Because the {\it Kepler} EBs have a wide range of
orbital periods,$P$, and are relatively faint, a comprehensive RV
survey is impractical.  Various groups are focusing on sub-samples,
such as transiting hot compact objects \citep{Coughlin11} or low-mass
stars \citep{Rowe10}. We are targeting $\sim$100 {\it Kepler} EBs with
light curves that predict low-mass secondaries.  Our objective is to
measure the dynamical masses of these stars to $3\%$ or better.
Binaries with mass ratio, $q=M_2/M_1$, much less than one, have a
secondary-to-primary flux ratio, $\alpha$, that is very small in
visible light, but more favorable in the infrared.  Such systems can
often be efficiently solved as SB2s by combining visible light
spectroscopy that precisely measures the primary RVs, with a smaller
number of infrared observations that detect the secondary
\citep[e.g.,][]{Bender08}.

We are observing our EBs with the H-band APOGEE spectrograph
\citep{Wilson10}, which is part of the Sloan Digital Sky Survey-III
\citep{Eisenstein11}.  APOGEE's multi-object capabilities can survey a
large number of EBs as SB2s, but provide a small number of epochs for
each (typically from three to six).  We are supplementing these
observations with visible light spectroscopy from the fiber-fed High
Resolution Spectrograph \citep[hereafter, HRS;][]{Tull98} on the 9.2m
Hobby-Eberly Telescope \citep[hereafter, HET;][]{Ramsey98}. The HET
queue-scheduled operation \citep{Shetrone07}, combined with the
precise $P$ measured by {\it Kepler}, provides efficient observations
of our EB sample at targeted orbital phases.  In most cases our HRS
spectra provide single-lined binary orbits (SB1s), but some, including
the results presented here, have the sensitivity to directly solve an
EB as an SB2.

Here, we present new observations of the circumbinary planet system
Kepler-16 as an SB2. Section~2 describes the Kepler-16 system and discusses the
importance of circumbinary planet systems.  Sections 3, 4, and 5,
present our observations, data analysis, and orbit fitting procedures.
Section 6 compares our results with D11 and draws some relevant
conclusions. Future papers will describe our
ongoing survey in more detail and present dynamical results from
additional EBs.

\section{Circumbinary planets \& the Kepler-16 system}
Planets orbiting binary stars, or circumbinary planets, likely provide
key insights towards understanding planet formation processes,
so discovering and characterizing these systems is a high priority.
However, such planets are not well suited to discovery with 
precision RV techniques that have yielded most of the confirmed planet
population because measuring precise velocities in a binary spectrum
is challenging.  Attempts to circumvent this problem by combining RVs
with interferometric observations have resulted in some of the most
precise stellar masses known, but have not yet detected planets
\citep{Konacki05b,Konacki10}.  Measuring transit timing variations of
EBs has become the technique of choice in circumbinary planet
searches. Eclipse timing variation signals, interpreted as due to a
planetary mass object, have been detected around HW Virginis, CM Dra,
SS Serpens, and HU Aq, although the planetary interpretation has also
been questioned \citep[e.g.,][]{Horner11,Wittenmyer12}.

Three circumbinary planets have recently been identified with
\emph{Kepler} photometry, and characterized through the application of
a three-body PDM \citep{Carter11}: Kepler-16b (D11), Kepler-34b, and
Kepler-35b \citep{Welsh12}.  The Kepler-16 system is particularly
interesting for constraining stellar mass-radius relationships because
its stellar components are both low-mass.  From the PDM, D11
determined that Kepler-16 has $q=0.2937\pm0.0006$ (J.\ Carter, private
communication).  When combined with single-lined TRES spectroscopy,
they derived masses of $M_A=0.6897 \pm 0.0035$ and $M_B=0.20255
\pm0.00065$, for Kepler 16-A \& B respectively.  Their spectroscopy
was unable to detect Kepler-16B, which has $\alpha=0.015$ in the
\emph{Kepler} bandpass.

Given the small, but significant, discrepancies between stellar radii
measured for low-mass stars and current stellar models, obtaining
robust absolute mass measurements for individual systems is critical
to anchor the absolute offset for the models.  The PDM
provides an estimate of stellar masses based on a three-body Newtonian
model, and is a relatively new and powerful technique that may be
applicable to numerous systems with precise photometry and multiple
transiting objects.  An important part of scientific investigation is
the independent confirmation of results and techniques, preferably
using different instruments and measurements of independent variables.
Observations of a binary as an SB2 use a Keplerian model (hereafter,
KM) to estimate $q$ and absolute masses, and are a well established
technique. The SB2 observations of Kepler-16 we present here directly
test the PDM results.

\section{Observations with the Hobby-Eberly Telescope} \label{observations}
We obtained six observations of Kepler-16 with the HRS, from September
through October 2011, using the 316g7940 cross-disperser, a 2$\arcsec$
diameter fiber that encompassed both stars, and a slit providing
R$\sim$30,000.  Each observation was a  1200 s integration,
which yielded S/N$\sim$300 per resolution element, and was bracketed
before and after with a ThAr hollow-cathode lamp exposure for
wavelength calibration. We used the same procedure on two known single
stars, HD17230 and GJ905, to serve as spectral templates for
Kepler-16A \& B, respectively.  HD17230 has spectral type K6V, with
$\mathrm{T_{eff}=4442\pm44 K}$, $\log g=4.89\pm0.06$, and
$\mathrm{[M/H]=0.02\pm0.03}$ \citep{Valenti05}.  GJ905 has spectral
type M5.5V, with $\mathrm{T_{eff}=2800\pm100 K}$, and
$\mathrm{[M/H]=0.00\pm0.25}$ \citep{Leggett00}.

We extract HRS spectra using a custom optimal extraction pipeline,
similar to that described by \citet{Cushing04}, written by us in the
Interactive Data Language (IDL).  This pipeline automatically performs
basic image processing, including overscan correction, bias
subtraction, and flat fielding.  It then automatically traces the
spectral orders in each HRS target observation, computes the optimal
fiber profile for each order, and carries out the extraction.  ThAr
calibration spectra are extracted in a similar manner, using a fiber
profile determined from exposures of twilight sky or a flat lamp
observed through the target fiber.  A multi-order solution equating
pixel position to wavelength is fit to the ThAr spectrum using the
linelist of \citet{Murphy07}, and then applied to the corresponding
target spectrum.  We found no significant difference between the two
ThAr frames that bracketed each observation, indicating that HRS was
stable throughout each integration, and so arbitrarily used the later
ThAr solution for each observation.  To facilitate use by other
investigators, we have placed our extracted Kepler-16 spectra in the
NASA Exoplanet
Archive\footnote{http://exoplanetarchive.ipac.caltech.edu}.

\def\arraystretch{1.1}

\begin{deluxetable*}{cccrr}
  \tablecaption{HRS Radial Velocity Measurements of Kepler-16A \& B\label{velocities}}
  \tablecolumns{5}
  \tablewidth{0in}
  \tablehead{
    \colhead{UT Date} & \colhead{BJD -- 2,400,000} & \colhead{Orbital Phase} & \colhead{\va (\kps)} &  \colhead{\vb (\kps)}}
  \startdata
2011 Sep 27    & 55831.677860 & 0.607 & $-39.175 \pm 0.026$ & -15.39 $\pm$ 0.80 \\
2011 Oct \phn6 & 55840.643682 & 0.825 & $-47.615 \pm 0.025$ & \phs13.32 $\pm$ 1.03 \\
2011 Oct \phn8 & 55842.631447 & 0.874 & $-46.510 \pm 0.030$ & \phs\phn8.88 $\pm$ 1.08 \\
2011 Oct 15    & 55849.627832 & 0.044 & $-30.288 \pm 0.023$ & -44.34 $\pm$ 0.63 \\
2011 Oct 20    & 55854.601732 & 0.165 & $-21.244 \pm 0.026$ & -75.20 $\pm$ 0.82 \\
2011 Oct 24    & 55858.588560 & 0.262 & $-20.917 \pm 0.025$ & -76.80 $\pm$ 0.78
\enddata
\end{deluxetable*}

We used a synthetic telluric absorption function generated with the
LBLRTM atmospheric model \citep{Clough05} to identify spectral regions
contaminated by telluric absorption, and excluded them from subsequent
analyzes.  We removed sharp night-sky OH emission features by linearly
interpolating over them.  In the 316g7940 configuration many orders
overlap off-blaze wavelengths.  We combined these orders into five
contiguous segments which were free of telluric absorption:
6180--6270\AA, 6340--6450\AA, 6600-6850\AA, 7440--7580\AA, and
8460--8870\AA. In each bandpass the Kepler-16 binary has a
different $\alpha$.  Using BT-Settl models \citep{Allard11} that
correspond to the $\mathrm{T_{eff}}$, $\log(g)$, and $\mathrm{[Fe/H]}$
derived by \citet{Winn11} and D11 (and shown in Fig 1) we estimate the
$\alpha$ in each bandpass to be 0.00522, 0.00743, 0.00780, 0.0228, and
0.0286, respectively.

\begin{figure}[bp]
\begin{center}
\includegraphics*[width=85mm]{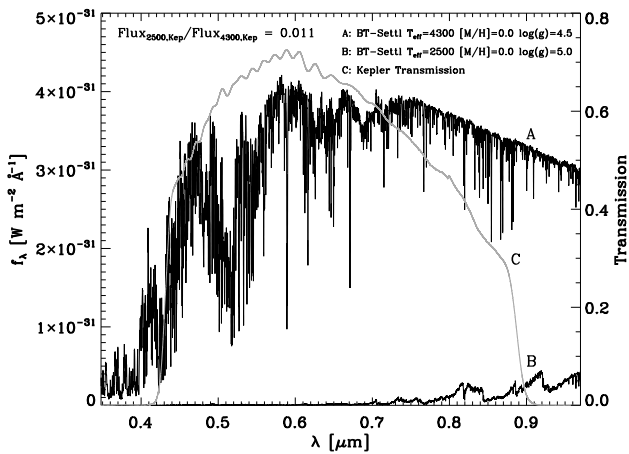}
\end{center}
\caption{BT-Settl models corresponding to Kepler-16A \& B, plotted
  along with the {\it Kepler} response function. The $\alpha$ of these
  models in the \emph{Kepler} bandpass, $\sim 1\%$, is consistent with
  the value measured by D11; in the HRS bandpasses $\alpha$ ranges
  from $\sim0.5\%$ to $\sim3\%$.} \label{fig:fluxcontrast}
\end{figure}

\section{SB2 Analysis}\label{sb2analysis}

The cross-correlation algorithm TODCOR \citep{Zucker94} simultaneously
cross-correlates two template spectra against a target spectrum
containing the blended light from a binary to disentangle the
component RVs.  We used TODCOR, along with the HD17230 and GJ905
templates, to measure the RVs of Kepler-16A \&
B in our HRS spectra.  Following \citet{Zucker03b}, we solved each bandpass
independently with $\alpha$ locked to the corresponding value
estimated in \S\ref{observations}, and combined the resulting
correlation surfaces with a maximum-likelihood analysis
\citep{Zucker03}.  TODCOR is more sensitive to the RV of each
component than to the $\alpha$, and in high-contrast binaries where
the flux from the primary dominates,  a
discrepancy of a factor of 2--3 between the chosen $\alpha$ and the
true value can still yield equivalent RVs.  Similarly, small mismatches in
the target and template metallicity primarily affect the ability of
TODCOR to derive an optimal $\alpha$, and do not affect the measured
RVs \citep{Bender05}.  By locking $\alpha$ across all HRS epochs, we
minimize the effect on the RVs of discrepancies with the BT Settl
$\alpha$ or template metallicity mismatch.

In analyzing each observation we restricted the secondary velocity
search to the range bounded by $q=1$ and the minimum-mass from the
mass-function, which are constrained by the SB1 solution and the
orbital phase, and chose the largest amplitude peak in this range.
The component RVs, \va{} and \vb{}, are measured by fitting the top
six to eight points of this peak with a quadratic.  Uncertainties are
derived from the maximum-likelihood formalism of \citet{Zucker03},
which accounts for spectral bandwidth, correlation peak sharpness, and
the spectral line S/N, but may exclude systematic uncertainties due to
the instrument or templates.  Our templates share many of the
same spectral features, so the M6V template correlates reasonably well
with the Kepler-16A spectrum and results in a spurious peak in
secondary velocity equal to \va.  In all observations except for 2011
Oct 15, the true secondary peak was well separated from the spurious
peak and could be measured directly.  For 2011 Oct 15, the secondary
peak was blended in the blue wing of the spurious peak.  We fit the
unblended portion of the spurious peak with a Gaussian and subtracted
it, which provided a clean secondary peak to measure \vb.

\begin{figure}[tp]
\begin{center}
\includegraphics*[width=85mm]{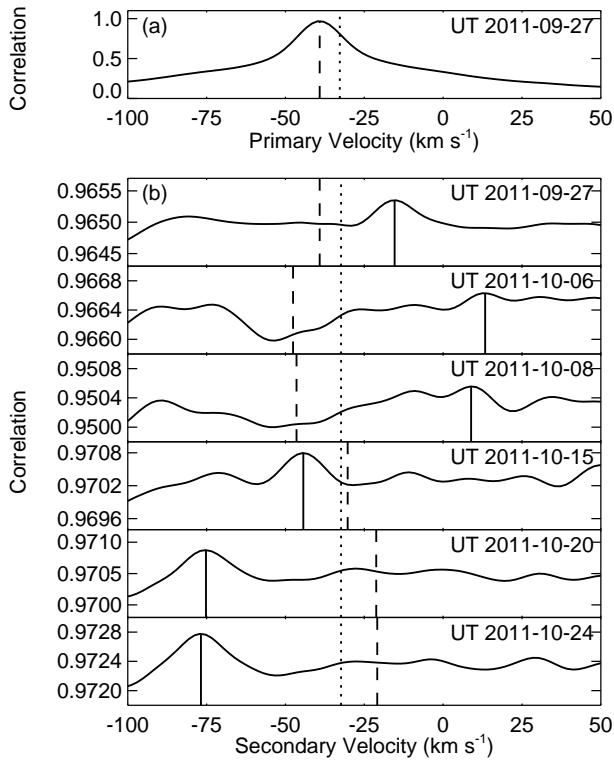}
\end{center}
\caption{Cuts through correlation surfaces vs.\ velocity, showing
  \emph{(a)} an example primary peak, and \emph{(b)} secondary peaks
  for each observation.  Primary cuts not shown have the same general
  shape as \emph{(a)}.  Vertical dashed and solid lines indicate the
  measured \va{} and \vb{}, respectively.  Dotted
  lines show the systemic velocity, $\gamma_A$.  The spurious secondary peak at
  \va=\vb{} (\S\ref{sb2analysis}) has been removed for clarity.} \label{fig:todcor}
\end{figure}

Table~\ref{velocities} lists the midpoint Barycentric Julian Date
(BJD), the corresponding orbital phase, and our measured RVs for each
HRS observation.  Figure~\ref{fig:todcor} shows cuts through the
correlation surfaces, along with the measured RVs.  We
are monitoring the HRS long-term stability by observing stars known to
be intrinsically stable to a few $\mathrm{m\,s^{-1}}$.  These
observations, taken over many months, show RV variations with
RMS$\sim$20$\,\mathrm{m\,s^{-1}}$, which is less than the \va{}
uncertainties reported in Table~\ref{velocities}.  The secondary peaks
for 2011 Oct 06 and 2011 Oct 08 are weaker than for the other dates
(Figure~\ref{fig:todcor}); this is likely due to lower intrinsic S/N,
and is reflected in the uncertainties reported in
Table~\ref{velocities}.

\section{Derivation of Orbital Parameters and Dynamical Stellar Masses}\label{orbitfitting}

Measuring Kepler-16 as an SB2 allows us to use a KM to
derive a dynamical mass-ratio, $q$, and dynamical masses, $M_A$ and
$M_B$, for the stellar binary.  These values constrain the stellar
mass-radius relationship and test the 23 parameter PDM used by D11.
For reference, Table~\ref{orbpars} lists the relevant orbital
parameters from D11.  For our purposes, $q$ is the most fundamental
parameter describing the EB, and can be derived without any orbit
modeling using the procedure of \citet{Wilson1941}.  The
``Wilson Test'' fits a straight line through a plot of \va{} versus
\vb{}: the negative slope of this line directly yields
$q=0.2986\pm0.0031$.

\begin{deluxetable}{lc}
  \tablecaption{Kepler-16 Orbital Parameters and Stellar Masses\label{orbpars}}
  \tablecolumns{2}
  \tablewidth{85mm}
  \tablehead{\colhead{Parameter} & \colhead{Value}}
  \startdata
  \multicolumn{2}{c}{\emph{Photometric-Dynamical Model (D11)}} \\ \\
  $P$ (days)   &  $41.077580\pm0.000008$ \\
  $T_t$ & $2454965.657623\pm0.000058$ \\
  $e$          & $0.15944^{+0.00061}_{-0.00062}$ \\
  $\omega$ (deg) &  $263.464^{+0.026}_{-0.027}$\\
  $i$ (deg)    &  $90.3401^{+0.0016}_{-0.0019}$\\
  $a$ (AU)     &  $0.22431^{+0.00035}_{-0.00034}$\\
  $q$ & $0.2937\pm0.0006$\\
  $M_{\mathrm{A}}$ (M$_\odot$) & $0.6897^{+0.0035}_{-0.0034}$\\
  $M_{\mathrm{B}}$ (M$_\odot$) & $0.20255^{+0.00066}_{-0.00065}$\\
  $R_{\mathrm{A}}$ (R$_\odot$) & $0.6489^{+0.0013}_{-0.0013}$\\
  $R_{\mathrm{B}}$ (R$_\odot$) & $0.22623^{+0.00059}_{-0.00053}$\\ \\
  \multicolumn{2}{c}{\emph{SB2 Spectroscopy (this work)}} \\ \\
  $e$ & $0.15894\pm0.00079$\\
  $\omega$ (deg) & $263.287\pm0.041$\\
  $K_A$ (\kps)   & 13.642 $\pm$ 0.010 \\
  $K_B$ (\kps)   & 45.56 $\pm$ 0.47 \\
  $\gamma_A$ (\kps) & -33.7551$\pm$ 0.0064\\
  $\gamma_B$ (\kps) & -33.32 $\pm$ 0.35 \\
  $q$            & 0.2994 $\pm$ 0.0031 \\ 
  $M_{\mathrm{A}}$ (M$_\odot$) & $0.654\pm0.017$\\
  $M_{\mathrm{B}}$ (M$_\odot$) & $0.1959\pm0.0031$
\enddata
\end{deluxetable}

The \emph{Kepler} photometry timing constrains the orbital period,
$P$, and the time of transit, $T_t$, with precision that cannot be
improved by new spectroscopy.  Table~\ref{orbpars} lists $P$ and $T_t$
from D11 that we use throughout our analysis\footnote{The D11
  Supporting Online Material incorrectly states that $T_t$ is relative
  to BJD 2455000; actually, the values are given relative to BJD
  2454900.}. We expand on the ``Wilson Test'' results by solving
Kepler's equations for the HRS RVs, while fixing $P$ and $T_t$ to
their \emph{a priori} values, yielding eccentricity, $e$, longitude of
periastron, $\omega$, semi-major axes, $K_A$ and $K_B$, and systemic
velocity $\gamma$.  We propagated the D11 uncertainty on $P$
throughout the analysis, but simplified the procedure by using the D11
ephemeris $T_t$ closest to our HRS observations and neglecting its
uncertainty; the impact on the final $q$, $M_A$, and $M_B$ should be
negligible.  D11 derived precise values for $e$, and $\omega$,
but because these are osculating parameters we chose to re-solve for
them.

The TRES RVs reported by D11 constrain the spectroscopic orbit of
Kepler-16A.  We used the Levenberg-Marquardt fitting code MPFIT
\citep{Markwardt08} to solve Kepler's equations for these RVs,
yielding a purely spectroscopic set of parameters from the D11
results: $e=0.15983\pm0.00085$, $\omega =
263.268^{\circ}\pm0.040^{\circ}$, $K_A=13.666\pm0.012$\kps,
$\gamma_A=-32.7765\pm0.0061$\kps, with reduced chi-squared of
$\chi_{\nu}^2=1.11$ and 15 degrees-of-freedom (DOF).  An independent
fit to our HRS primary RVs gives $e=0.1602\pm0.0020$, $\omega =
263.46^{\circ}\pm0.12^{\circ}$, $K_A=13.598\pm0.022$\kps, and
$\gamma_A=-33.770\pm0.018$\kps, with $\chi_{\nu}^2=1.58$ for 2 DOF.
The HRS and TRES reference frames are offset by $\sim$ 1\kps, which
can be accounted for with an offset parameter; otherwise, the two
solutions are in agreement, considering the paucity of HRS RVs.

We then simultaneously fit the HRS and TRES primary RVs and the HRS
secondary RVs, accounting for the reference frame offset.  The number
of HRS secondary RVs is about four times fewer than the combined
number of primary RVs, and they have much lower precision, so they do
not further constrain $e$ and $\omega$.  Table~\ref{orbpars} lists the
resulting orbital parameters and mass ratio, $q=0.2994\pm0.0031$,
which is completely consistent with $q$ derived from the ``Wilson
Test'', and Fig~\ref{fig:sb2plot} shows the corresponding SB2 RV curves.
This solution has $\chi_{\nu}^2=1.013$, with 25 DOF. Our velocity
precision is sufficient to require different systemic velocities,
$\gamma_A$ and $\gamma_B$, for each star, due to the combined
convective blue-shift and gravitational redshift of the two stars and
templates \citep{Pourbaix02, Dravins99}.  We estimate the
combined effect at $\mathrm{\sim400\,m\,s^{-1}}$, but did not attempt
to disentangle the individual contributions.  The uncertainty measured
for $\gamma_B$ implies that $\gamma_A$ and $\gamma_B$ differ by only
$\sim$1.25$\sigma$.

To examine the influence of systematic differences between the
TRES and HRS RVs, we re-computed the orbital parameters using only the
HRS data.  As before, $e$ and $\omega$ are dominated by the primary
RVs.  This yields $e$, $\omega$, $K_A$, and $\gamma_A$ that are
numerically equivalent to the HRS SB1 derived above, and $K_B$ and
$\gamma_B$ that are equivalent to the values in Table~\ref{orbpars},
with $\chi_{\nu}^2=1.12$ for 6 DOF.  The resulting $q=0.2986\pm0.0032$
is identical to that derived from the ``Wilson Test'', and
demonstrates that nearly all of the reported uncertainty is
contributed by the secondary RVs: including or excluding the TRES RVs
does not meaningfully change our derived masses or their precision.

Our measured $K_A$ and $K_B$ yield $q=K_A/K_B=0.2994\pm0.0031$ and,
using the physical constants suggested by \citet{Harmanec11},
$M_A\sin^3{i}=0.654\pm0.017\msun$ and
$M_B\sin^3{i}=0.1959\pm0.0031\msun$. When combined with the
inclination measured by D11, we can solve directly for the dynamical
masses of Kepler-16: $M_A=0.654\pm0.017\msun$ and
$M_B=0.1959\pm0.0031\msun$.

\section{Discussion}\label{discussion}

\begin{figure}[tbp]
\begin{center}
\includegraphics*[width=90mm]{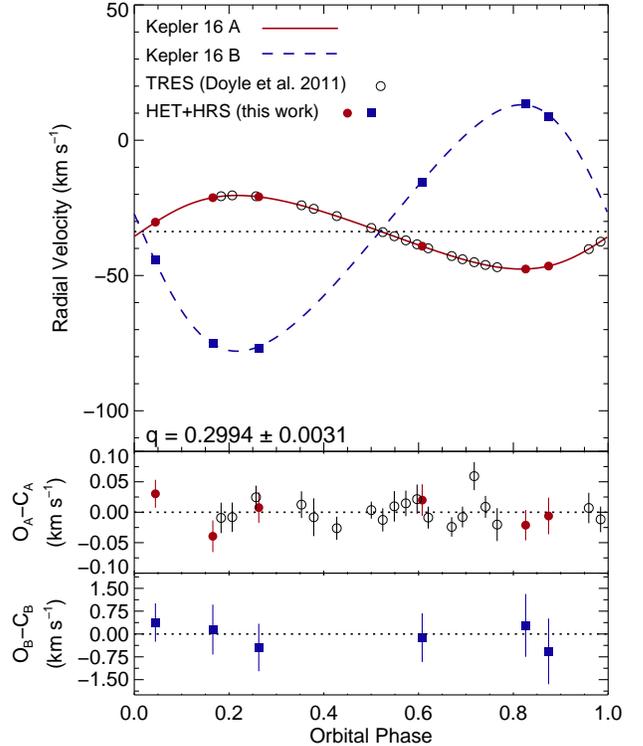}
\end{center}
\caption{RV vs.\ orbital phase for Kepler-16A \& B.  \emph{Upper
    Panel:} Solid circles and squares show our HRS \va{} and \vb{},
  respectively.  Open circles show the D11 TRES \va{}, offset to the
  HRS RV frame. The solid and dashed lines show the best orbital
  solution (Table~2).  \emph{Lower Panels:} Observed -- calculated
  residuals for the HRS and D11 RVs.} \label{fig:sb2plot}
\end{figure}

We have measured $q$ to $\sim1\%$, consistent with D11 at
$\sim2\sigma$, and $M_B$ to $\sim1.5\%$.  We performed several tests
to explore the $q$ discrepancy and the robustness of our derived orbital
parameters.  Using the D11 orbit, we calculate predicted
$K_A=13.661\pm0.030$\kps{} and $K_B=46.512\pm0.076$\kps.  The
difference between this $K_A$ and our measured value reported in
Table~\ref{orbpars} cannot account for the $q$ discrepancy.  The
difference between the predicted $K_B$ and our measured value is
2$\sigma$. An error in our measured $K_B$ of $\sim1\kps$, consistent
with our reported uncertainty at 2$\sigma$, would account for the
discrepancy.  However, the $\mathrm{O_B-C_B}$ residuals
(Fig~\ref{fig:sb2plot}) suggest that our \vb{} uncertainties are
overestimated. 

Due to the small number of \vb{} measurements, a systematic error in
one might skew our measured $K_B$.  To test this, we repeated the KM
fit six times, excluding each \vb{} in turn.  This produced a range in
$K_B$ from $45.36\pm0.59\kps$ to $45.71\pm0.61\kps$, corresponding to
$q$ from $0.2984\pm0.0040$ to $0.3007\pm0.0039$, and $\gamma_B$ from
$-33.21\pm0.40$ to $-33.46\pm0.39$.  These distributions are symmetric
around the values in Table~\ref{orbpars}, and confirm that no single
HRS SB2 observation is skewing the measured $q$.

We considered that our analysis of the SB2 RVs might be affected by
the planet's orbital motion.  However, this is unlikely because
Kepler-16b has a nearly circular orbit with $P\sim$229 days (D11),
while our HRS observations were obtained over just 28 days.  None of
our HRS observations coincided with a transit event, which would
subject the RVs to a Rossiter-McLaughlin shift \citep{Winn11}.
Sunlight scattered off the moon might confuse our SB2 analysis, but
the moon was down for three observations and at least $70^{\circ}$
from Kepler-16 for the other three.  No scattered light was detected
in HRS sky fibers exposed coincident with our observations, so
moonlight contamination is unlikely.

We have used dynamical SB2 measurements to derive $q$, $M_A$, and
$M_B$ for the Kepler-16 EB, and have confirmed the $q$ reported by D11
from their PDM to $\sim$2\%.  In addition, the $e$ and $\omega$
measured by the two techniques are equivalent.  The claimed precision
($q$ $\sim1\%$ by us and $\sim0.2\%$ by D11), results in a
$\sim$2$\sigma$ disagreement that we cannot account for.  It is
possible that our measurement errors are underestimated due to an
unidentified systematic effect, or that the D11 dynamical analysis is
erroneous, potentially due to the presence of additional bodies in the
system.  However, the latter is difficult to understand in light of
the continued accuracy of transit and eclipse ephemerides in
subsequent \emph{Kepler} quarters of data (J.\ Carter, private
communication).  In addition to verifying the PDM result, we have
demonstrated that masses with 1--2$\%$ precision can be measured for
\emph{Kepler} EBs, even when no planet transits are present.

\acknowledgments We thank the referees, J.~Carter, G.~Torres, and
L.~Doyle, for a thorough report, and J.~Carter for useful discussions
about the D11 result prior to our initial submission of the
manuscript.  This work was partially supported by the Center for
Exoplanets and Habitable Worlds, which is supported by the
Pennsylvania State University, the Eberly College of Science, and the
Pennsylvania Space Grant Consortium.  We acknowledge support from the
NAI, PSARC, and NSF grant AST-1006676. Data presented herein were
obtained at the Hobby-Eberly Telescope (HET), a joint project of the
University of Texas at Austin, the Pennsylvania State University,
Stanford University, Ludwig-Maximilians-Universit\"{a}t M\"{u}nchen,
and Georg-August-Universit\"{a}t G\"{o}ttingen. The HET is named in
honor of its principal benefactors, William P. Hobby and Robert
E. Eberly.

\end{document}